\documentclass[12pt, draftclsnofoot, onecolumn]{IEEEtran}

\makeatletter
\def\endthebibliography{%
	\def\@noitemerr{\@latex@warning{Empty `thebibliography' environment}}%
	\endlist
}
\makeatother

\usepackage{ifpdf}

\usepackage{cite}

\usepackage{graphicx}
\usepackage{epstopdf} 
\usepackage{xcolor}
\ifCLASSINFOpdf

\fi

\usepackage[cmex10]{amsmath}
\usepackage{amsfonts,amssymb}

\usepackage{algorithmic}

\usepackage{array}

\ifCLASSOPTIONcompsoc
\usepackage[caption=false,font=normalsize,labelfont=sf,textfont=sf]{subfig}
\else
\usepackage[caption=false,font=footnotesize]{subfig}
\fi

\usepackage{dblfloatfix}

\usepackage{url}

\usepackage{arydshln}
\usepackage{enumerate}

\usepackage[mathcal]{euscript}

\newtheorem{lemma}{Lemma}

\newtheorem{remark}{\textbf{Remark}}
\newcommand{\vect}[1]{\mathbf{#1}}

\newcommand{\bs}[1]{\boldsymbol{#1}}
\newcommand{\erfc}{\text{erfc}}

\def\beq{\begin{equation}}
\def\eeq{\end{equation}}

\graphicspath {{Figures/}}

\begin{document}
	
	\bstctlcite{IEEEexample:BSTcontrol}
%
\title{Grant-Free  Massive MTC-Enabled Massive MIMO: A Compressive Sensing Approach}
%
%
%

	\author{Kamil~Senel,~\IEEEmembership{Member,~IEEE,}
	and~Erik~G.~Larsson~\IEEEmembership{Fellow,~IEEE}
		\thanks{Parts of this work were  presented at the IEEE Global Communications Conference (GLOBECOM) 2017 \cite{senel2017device}, and at the International Workshop on Smart Antennas (WSA) 2018 \cite{senel2017mMTCwsa}.

			The authors are with the Department of Electrical Engineering (ISY), Link\"{o}ping University, 581 83 Link\"oping, Sweden.}\thanks{This work was supported
			by the Swedish Research Council (VR) and ELLIIT.}
	}

%
%



\maketitle

\begin{abstract}
A key challenge of  massive MTC (mMTC), is the joint detection of   device activity and decoding
of data. The sparse characteristics of mMTC makes compressed sensing (CS) approaches a promising solution to the device detection problem. However, utilizing CS-based approaches for device detection along with channel estimation, and using the acquired estimates for coherent data transmission is suboptimal, especially when  the goal is to convey only a few bits of data. 

First, we focus on the coherent transmission and demonstrate that it is possible to obtain more accurate channel state information by combining  conventional estimators with CS-based techniques. Moreover, we illustrate that even simple power control techniques can enhance the device detection performance in mMTC setups. 

Second,  we devise a new  non-coherent transmission scheme for mMTC and specifically for  grant-free random access. We  design an algorithm that jointly detects device activity along with embedded information bits. The  approach leverages elements from  the approximate message passing (AMP) algorithm, and exploits the structured sparsity introduced by the non-coherent transmission scheme. Our analysis reveals that the proposed approach
has superior performance compared to application of the original AMP approach. 
\end{abstract}


%
\IEEEpeerreviewmaketitle

\section{Introduction} \label{sec:Intro}

\IEEEPARstart{M}{achine}-type-communication (MTC) compels a paradigm shift in wireless
communication due to the diverse data traffic characteristics and
requirements on delay, reliability, energy consumption, and
security.  A key scenario of MTC, referred as massive MTC (mMTC),
corresponds to providing wireless connectivity to a massive number of
low-complexity, low-power machine-type devices \cite{deliverable2015d6}. These devices
enable  various emerging smart
services in the fields of  healthcare, security,
manufacturing, utilities and transportation
\cite{ericsson}. 

Cellular networks are a potential candidate
to accommodate the emerging MTC traffic thanks to the existing
infrastructure and wide-area coverage
\cite{centenaro2016long}. However, previous generations of
cellular systems are designed for human-type communication (HTC)
which aims for high data rates using large packet sizes
\cite{bockelmann2016massive}. The integration of MTC along with HTC in
cellular networks requires the handling of diverse communication
characteristics. Moreover, unlike HTC, in MTC the data traffic is uplink-driven
with packet sizes going down as low as a few bits
\cite{boccardi2014five}. An  example of a \emph{single-bit} transmission
is the transmission of ACK/NACK bits \cite{larsson2012piggybacking}. In the mMTC context, the amount of signaling overhead per packet can  become very
significant compared to   traditional setups with mainly human-driven traffic \cite{dawy2017toward}. 

\begin{figure}[tb]
	\begin{center}
		\includegraphics[trim=0cm 0cm 0cm 0cm,clip=true, scale = .7]{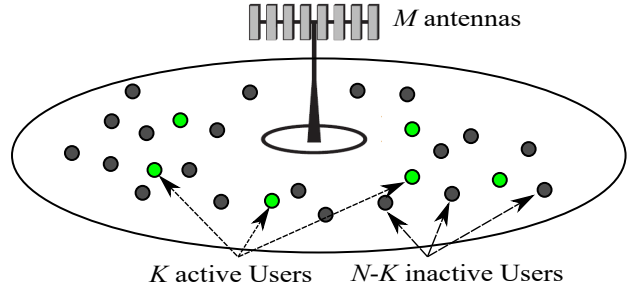}

		\caption{mMTC scenario: An $M$-antenna base station serves $N$ users, of which $K$ are active at a given point in time.}
		\label{fig:SystemSetup} 
	\end{center}
\end{figure}

In mMTC, only a small fraction of the devices is active at a time. One reason for this sporadic traffic pattern is the inherent intermittency of the traffic (especially for sensor data), but the use
of higher-level protocols that generate bursty traffic also contributes. 
The setup of interest is depicted in Fig. \ref{fig:SystemSetup}. Here, a base station (BS) with $M$ antennas provides service to $N$ devices and
among these $N$ devices, only $K$ are active at 
a given time. Our focus will be on systems with Massive MIMO technology
such that $M$ is large. Massive MIMO
is an important component of the 5G physical layer,
as it enables the multiplexing of many devices in the same time-frequency resources as well as a range extension owing to the coherent beamforming gain \cite{redbook}.

The intermittency of mMTC traffic calls for efficient mechanisms for random access.
Here we focus on \emph{grant-free random access}, where devices access the network
without a prior scheduling assignment or a grant to transmit. Owing to
the massive number of devices, it is  impossible to assign orthogonal pilot
sequences to every device.
This inevitably leads to collisions between the devices. Conventionally,
such collisions are handled through collision resolution mechanisms \cite{sesia2011lte, de2017randomM}. 
Standard ALOHA-based  approaches are not suitable
for mMTC, as ALOHA suffers from low performance when the number of
accessing devices is large \cite{boljanovic2017user}. A promising class
of collision resolution methods, known as compressed sensing (CS)
techniques, have been considered for device detection in mMTC
\cite{choi2017compressed}. With that approach, all active users transmit their unique
identifiers concurrently, and the base station (BS) detects the set of active devices
based on the received signal. Moreover, unique user identifiers can be
utilized as a sensing matrix to estimate the channels along with
the device detection \cite{nan2015efficient}. The CS algorithms are shown to
outperform  conventional channel estimation techniques when the
device activity detection is to be performed jointly with channel estimation \cite{choi2015downlink}. However, conventional channel estimation
techniques may also be employed once CS-based device detection has been
accomplished. Under the assumption that perfect channel state
information (CSI) is available, the channel states can be utilized as
a sensing matrix and the joint active device and data detection
problem can be tackled by CS-based techniques both for single-antenna \cite{du2017efficient,wang2015compressive} and MIMO setups
\cite{gao2016compressive, garcia2015low}.

In coherent transmission, the detection of active devices and the estimation of their channels
is followed by payload data transmission. Coherent transmission in an mMTC setup has been investigated in \cite{de2017random} which proposes an approach that relies on pilot-hopping over multiple coherence intervals. A paper that investigates the spectral efficiency of a CS-based approach for mMTC setup is \cite{liu2017massive}.  
However, the
acquisition of accurate channel state information 
is a challenging task, which prompted researchers to
consider the possibility of non-coherent transmission schemes
\cite{liu2017novel,jing2016design}. Especially, for mMTC
where devices usually transmit small packets intermittently, using
resources to obtain CSI for coherent transmission may not be optimal.  

In this work, we consider the uplink transmission between a large
number of devices and a Massive MIMO BS. The BS aims to detect the set of
active devices and estimate their channels and decode a small amount
of data transmitted by the active devices. The approaches in the
literature employ coherent transmission based on  estimates
acquired from the CS-based algorithms. We demonstrate that the
minimum-mean square estimator, combined with CS-based techniques, 
can be utilized to obtain more accurate
CSI. Furthermore, a novel non-coherent transmission technique is introduced. A comparison
between coherent and non-coherent approaches reveals that non-coherent
transmission can significantly outperform coherent transmission  
in mMTC setups. Comparisons of coherent and non-coherent transmission techniques in multiple-antenna setups are available in the literature \cite{zheng2002communication}. It is known that generally, non-coherent transmission
outperforms coherent transmission. 
In this work, we provide a comparison under an mMTC setup with specific
focus on the  challenges that arise when  joint  device detection, channel estimation and data decoding must be performed with non-orthogonal pilots. 

 The specific contributions of our work are as follows: 
\begin{itemize}
	\item An analysis of the AMP algorithm demonstrates that the gains from increasing the number of BS antennas is comparable to increasing pilot sequence length, making  massive MIMO   a key enabler for MTC applications. (Section \ref{sec:DevDetAMP})  
		\item We investigate the effect of employing a power control approach suitable for mMTC setups, on device detection performance. The analysis reveals that power control provides significant improvement in terms of device detection. (Section \ref{sec:PowerControl})   
		
	\item We present a scheme which combines conventional channel estimation techniques with CS-based device detection algorithms, and derive a closed-form expression for the resulting achievable spectral efficiency. The proposed scheme significantly enhances the spectral efficiency for coherent transmission.  (Section \ref{sec:ChEstimation})   
	
		\item We introduce a novel non-coherent data transmission technique based on embedding information bits to the pilot sequences to be decoded during the user activity detection process. (Section \ref{sec:non-coh}) 

	\item We devise a new receiver
	based on approximate message passing that detects which devices are active, and detects their associated information bits, without using any prior information neither on the channel response nor on the user activity. (Section \ref{sec:AlgDes})
	
	\item We provide an extensive comparison between coherent and non-coherent transmission techniques and demonstrate that under mMTC setups, non-coherent transmission is more suitable for conveying small numbers of information bits. (Section \ref{sec:CohVsNon}) 
\end{itemize}
The paper in hand goes beyond our previous conference papers \cite{senel2017device,senel2017mMTCwsa}, by considering
 power control, non-coherent transmission for multiple bits,   detailing a  new modified AMP algorithm for the multi-bit case, and providing   several new experimental results and comparisons. Moreover, the analysis is carried out utilizing a novel receiver, which is designed for the proposed non-coherent scheme and provides additional performance gains compared to the original AMP algorithm.
 
\section{System Setup}\label{sec:SystemSetup}

We consider the uplink communication between a single base station with $M$ antennas and $N$ single antenna devices. Non-line of sight communication is assumed and the channel between device $n$ and the BS is modeled as 
\begin{equation}\label{eq:Ch}
\vect{g}_n = \sqrt{\beta_n}\vect{h}_n,~~\forall n = 1, \ldots,N, 
\end{equation}
where $\beta_n$ is the large-scale fading and $\vect{h}_n$ denotes the small-scale fading. The elements of $\vect{h}_n$ are assumed to be i.i.d. $\mathsf{CN}(0,1)$. The channel is constant and frequency-flat for $\tau$ samples called coherence interval (CI). 
The large-scale fading coefficients are assumed to be known at the BS and identical across antennas whereas the small-scale fading coefficients which change independently between CIs, are to be estimated in each CI. 
 
During coherent transmission, each CI is utilized for both channel estimation and data transmission, i.e., each active device transmit $\tau_p$-length pilot sequences and the remaining $\tau - \tau_p$ symbols are utilized for data transmission. In order to accomplish coherent data transmission, BS must detect the active devices, estimate their channels, and decode the transmitted data based on the acquired channel estimates. In traditional networks, an orthogonal pilot sequence is assigned to each device which requires pilot sequences of length $\tau_p \geq N$. Such an approach is not feasible for mMTC systems as the number of devices is large. Therefore, we consider a setup with non-orthogonal pilot sequences which are generated by sampling an i.i.d. symmetric Bernoulli distribution. Let $\sqrt{\tau_p}\bs{\varphi}_n$ denote the pilot sequence of the $n$th device with $\bs{\varphi}_n \triangleq [\varphi_{1,n}, \ldots, \varphi_{\tau_p,n} ]^T \in \mathbb{C}^{\tau_p \times 1}$ where $\varphi_{l,n} = (\pm 1 \pm j)/\sqrt{2\tau_p}$ and $\|\bs{\varphi}_n\|^2 = 1$. As a result of the Bernoulli distribution assumption, there are a finite number of unique pilot sequences and hence the probability that two devices have identical pilot sequences (called the ``collision probability'' here) is non-zero. If the sequences were generated by sampling an i.i.d. symmetric Gaussian distribution, the collision probability would be zero. However, as will be demonstrated later, pilot sequences based on Bernoulli distribution provide  better performance.
Let $\text{Pr}^{\mathrm{C}} (\tau_p,N)$ be the collision probability for a given number of devices, $N$, and a pilot sequence length, $\tau_p$. Then,
\begin{equation}
\text{Pr}^{\mathrm{C}} (\tau_p,N) = \begin{cases}1 - \prod\limits_{k = 1}^{N - 1 }\left(1 - \frac{k}{2^{2\tau_p}}\right), & N \leq 2^{2\tau_p},\\ 1, & N > 2^{2\tau_p}.\end{cases}
\end{equation}
In practice, the collision probability is negligible, for example,   with $N = 200$ devices and pilot sequences of length $\tau_p = 20$, the collision probability is $\sim 10^{-8}$. 

In our setup, we assume that the pilot sequences associated with each device are known at the BS. 
The justification is that in practice the BS would have  a list of devices that are associated with it, and their unique identifiers. The pilot sequences may then be created by a pseudo-random generator that uses the unique identifiers of the devices as seeds. Since these unique identifiers are known to the BS, the pilot sequence matrix is also known at the BS.
 Note that all
devices are not necessarily active in each of the coherence intervals;
only when they have data to transmit, they will communicate with the BS.  

The BS detects active devices in a given CI based on the received composite signal, $\vect{Y} \in \mathbb{C}^{\tau_p \times M}$ which is defined as
\begin{equation}\label{eq:recComp}
\vect{Y} = \sum_{n=1}^N \sqrt{\tau_p\rho_{ul}}\alpha_{n} \bs{\varphi}_{n}\vect{g}_{n}^T + \vect{Z}, 
\end{equation}  
where $\alpha_{n}$ is the device activity indicator for device $n$ with
$\text{Pr}(\alpha_n = 1) = \epsilon$ and $\text{Pr}(\alpha_n = 0) = 1 - \epsilon$; $\vect{Z}$ is additive white Gaussian noise with i.i.d. elements $\sim \mathit{CN}(0, \sigma^2)$. The transmission power is denoted by $\rho_{ul}$ and it is identical for each device. In Section \ref{sec:PowerControl}, we investigate the performance when power control is employed. 

 Let $\bs{\Phi} = [\bs{\varphi}_1, \ldots, \bs{\varphi}_N] \in \mathbb{C}^{\tau_p \times N}$ be the pilot matrix and $\vect{X} = [\vect{x}_1, \ldots, \vect{x}_N]^H \in \mathbb{C}^{N \times M}$ be the effective channel matrix where
\begin{equation}
\vect{x}_n = \alpha_n \vect{g}_n.
\end{equation} 
Then, \eqref{eq:recComp} can be rewritten in vector notation as
\begin{equation} \label{eq:recCompMatrix}
\vect{Y} = \sqrt{ \tau_p\rho_{ul}} \bs{\Phi}\vect{X} + \vect{Z}.
\end{equation}
Note that, $\vect{X}$ has a sparse structure as the rows corresponding to inactive users are zero. The activity detection problem reduces to finding the non-zero rows of $\vect{X}$. 

The motivation of this work is based on finding efficient communication 
techniques for grant-free random access with small amounts of data in mobile systems. 
Conventional techniques that rely on channel estimates and employ coherent transmission may not be suitable for mMTC for two critical reasons. First, the coherence interval length, the duration in which the channel can be   assumed to be flat, limits the number of orthogonal pilots which in turn makes it harder to obtain accurate channel estimates. Second, allocating orthogonal pilots to each device is suboptimal, if possible at all, due to the intermittent nature of mMTC. Furthermore, utilization of higher frequency bands and relatively high mobility of devices in some mMTC scenarios, e.g.\ vehicular sensing, the coherence interval length is substantially smaller which compels different approaches for data transmission.

\section{Review of Approximate Message Passing}

The problem of detecting active devices is equivalent to finding the non-zero rows of $\vect{X}$ based on the noisy observations, $\vect{Y}$ and known pilot sequences, $\bs{\Phi}$. This problem can be modeled as a compressive sensing problem, as $\vect{X}$ has a row-wise sparse structure. For the single antenna setup, the problem reduces to the single measurement vector (SMV) reconstruction problem whereas with multiple antennas it becomes a multiple measurement vector (MMV) reconstruction problem. CS-based techniques are shown to outperform linear minimum mean square error (LMMSE) estimators in terms of device detection performance in various works \cite{choi2017compressed,choi2015downlink}. In this work, a low complexity CS algorithm called approximate message passing (AMP) \cite{kim2011belief,ziniel2013efficient} is utilized to recover the sparse $\vect{X}$. Next, we provide a brief review of the AMP algorithm. 

 Let $t$ denote the index of the iterations and let $\hat{\vect{X}}^t = [\hat{\vect{x}}^t_1, \ldots, \hat{\vect{x}}^t_N]^H$ be the estimate of $\vect{X}$ at iteration $t$. Then, the AMP algorithm can be described as follows:
\begin{IEEEeqnarray}{lll}
\hat{\vect{x}}^{t+1}_n &=& \eta_{t,n} \left( (\vect{R}^t)^H \bs{\varphi}_n + \hat{\vect{x}}^{t}_n \right) \label{eq:AMP-1}\\
\vect{R}^{t+1} &=& \vect{Y} - \bs{\Phi}\hat{\vect{X}}^{t+1} + \frac{N}{\tau_p} \vect{R}^t \sum_{n=1}^N \frac{\eta_{t,n}' \left( (\vect{R}^t)^H \bs{\varphi}_n + \hat{\vect{x}}^{t}_n \right)}{N}  \label{eq:AMP-2}
\end{IEEEeqnarray}
where $\eta(.)$ is a denoising function, $\eta(.)'$ is the first order derivative of $\eta(.)$ and $\vect{R}^t$ is the residual at iteration $t$ \cite{rangan2016vector}. The residual in \eqref{eq:AMP-2} is updated with a crucial term containing $\eta(.)'$, called the Onsager term, which has been shown to substantially improve the performance of the iterative algorithm \cite{donoho2009message}. 

An important property of AMP is that in the asymptotic region, i.e., as $\tau_p,~K,~N \rightarrow \infty$ while their ratios are fixed, the behavior is described by a set of state evolution equations \cite{rangan2011generalized}. In vector form, the state evolution is given by \cite{bayati2011dynamics}
\begin{equation}\label{eq:sigmaUpdate}
\vect{\Sigma}^{t+1} = \frac{\sigma^2}{\rho_{ul}\tau_p}\vect{I} + \frac{N}{\tau_p} \mathbb{E}\{ \vect{e}\vect{e}^H \}
\end{equation}    
where $\vect{e} = \eta (\vect{x}_{\beta} - (\vect{\Sigma}^t)^{\frac{1}{2}}\vect{w})-\vect{x}_{\beta}$; $\vect{w}\in \mathbb{C}^{M\times1}$ is a complex Gaussian vector with unit variance and $\vect{x}_{\beta} \in \mathbb{C}^{M\times1}$ has the distribution 
\begin{equation} \label{eq:distH}
p_{\vect{x}_{\beta}} = (1 - \epsilon)\delta + \epsilon p_{\vect{h}_{\beta}}.
\end{equation}
Here, $p_{\vect{h}_{\beta}} \sim \mathit{CN}(0, \beta \vect{I})$ is the distribution of the channel vector of the active device and $\delta$ is the dirac Delta at zero corresponding to the inactive device channel distribution. The expectation in \eqref{eq:sigmaUpdate} is taken with respect to $\beta$ and allows   performance analysis of the AMP algorithm as the update given by   \eqref{eq:AMP-1}-\eqref{eq:AMP-2} are statistically equivalent to applying a denoiser to the following \cite{rangan2011generalized}:
\begin{equation} \label{eq:equaivalentAMP}
\hat{\vect{x}}^t_n = \vect{x}_n + (\vect{\Sigma}^t)^{\frac{1}{2}}\vect{w} = \alpha_n \vect{h}_n +  (\vect{\Sigma}^t)^{\frac{1}{2}}\vect{w},
\end{equation}       
which decouples the estimation process for different devices.
The state evolution is shown to be valid for a wide range of Lipschitz continuous functions \cite{bayati2011dynamics}. For the multiuser detection problem, the following denoising function is used: 
\begin{equation}\label{eq:etaMMSE}
\eta_{t,n}(\hat{\vect{x}}_n^t) = v(\hat{\vect{x}}_n^t; \vect{\Sigma}^t )\beta_n\left(\beta_n \vect{I} + \vect{\Sigma}^t\right)^{-1} \hat{\vect{x}}_n^t 
\end{equation}   
where 
\begin{eqnarray}
v(\hat{\vect{x}}_n; \vect{\Sigma} ) &=& \frac{1}{1+\frac{1-\epsilon}{\epsilon}\text{det}(\vect{I} + \beta_n \vect{\Sigma}^{-1}) q(\hat{\vect{x}}_n; \vect{\Sigma} )}, \\
q(\hat{\vect{x}}_n; \vect{\Sigma} ) &=& \text{exp}\left(- \hat{\vect{x}}_n^H\left(\vect{\Sigma}^{-1}- (\vect{\Sigma} + \beta_n\vect{I})^{-1} \right)\hat{\vect{x}}_n \right).
\end{eqnarray}
The denoising function \eqref{eq:etaMMSE} is shown to be the MMSE for the equivalent system described by \eqref{eq:equaivalentAMP} in \cite{kim2011belief}. Notice that, when the active device are to be detected the MMSE given by \eqref{eq:etaMMSE}, is non-linear. 

Note that $v(\cdot)$ is a thresholding function based on the likelihood ratio which can be computed by considering two cases in \eqref{eq:equaivalentAMP}, device $n$ is active, i.e., $\alpha_n = 1$ and $\alpha_n = 0$ when it is inactive. For the case when $\epsilon = 1$, i.e., every device is active, \eqref{eq:etaMMSE} reduces to the linear MMSE estimator.

\begin{remark}
	State evolution provides an important   tool to analyze AMP. However, the equations defined in \eqref{eq:equaivalentAMP}, which decouple the estimation process for different devices, are only valid in the asymptotic region. More detail on the behavior of  AMP in the asymptotic region is given in Section \ref{sec:Asymp}.
\end{remark}

   \begin{table}[t!]
	\centering
	\caption{Simulation Parameters}
	\label{tbl:SysParameters}
	\begin{tabular}{l|l}
		\hline
		\textbf{Parameter} & \textbf{Value} \\ \hline
		Path and penetration loss at distance $d$ (km) & 130 + 37.6 $\log_{10}(d)$ \\ 
		Bandwidth ($B_w$)                 & 20 MHz         \\
		Cell edge length           & 250 m          \\
		Minimum distance           & 25 m           \\ 
		Total noise power ($\sigma^2$)         & 2$\cdot10^{-13}$ W \\ 
		UL transmission power ($\rho_{ul}$)         & $0.1$ W         \\  
		\hline
	\end{tabular}
	\vspace{-0.4cm}
\end{table}

\subsection{Device Activity Detection via AMP}\label{sec:DevDetAMP}

The AMP approach heavily relies on the sparsity in the device activity pattern. The so-called ''sparsity-undersampling tradeoff'' states that as sparsity decreases, the length of the pilot sequences must increase in order to achieve the same performance \cite{donoho2009message}. For the noiseless case, a lower bound on the length of pilot sequences for perfect recovery is given by $\tau_p \geq K$ \cite{kim2011belief}. The device detection problem has a key difference compared to the reconstruction problem: It is not necessary to reconstruct the signal perfectly, only the devices that transmit their pilot sequences must be detected. However, being able to detect devices without recovery does not render the reconstruction of $\vect{X}$ an unnecessary task, as the reconstruction process corresponds to the estimation of the channels, which will be investigated in Section \ref{sec:ChEstimation}. 

\begin{figure}[t!]
	\begin{center}
		\includegraphics[trim=.3cm 0cm 0cm 0.4cm,clip=true,width = 9cm]{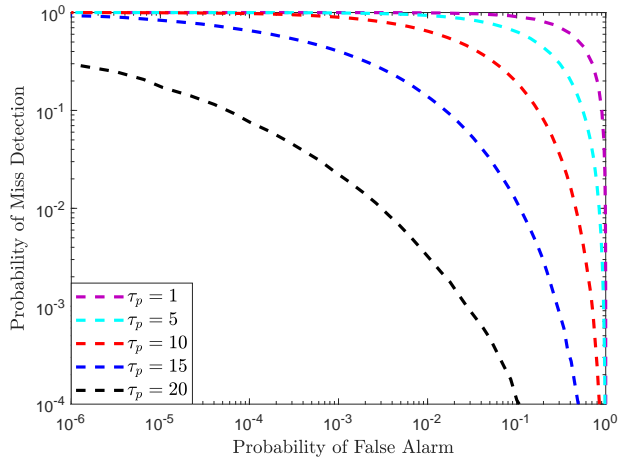}
		\caption{Probabilities of miss and false alarm for different  pilot sequence lengths, $\tau_p$, for $N = 200$ devices with a device access probability $\epsilon = 0.05$ and $M = 20$ antennas at the base station.}
		\label{fig:fig1}
	\end{center}
\end{figure}

 Fig.~\ref{fig:fig1} demonstrates the performance of the AMP algorithm for various pilot sequence lengths under a setup with $M = 20$, $N = 200$ and $\epsilon = 0.05$. The results illustrate that the performance highly depends on the pilot sequence length. As the pilot sequence length increases, the average correlation between pilot sequences of different devices decreases. Note that the improvement is especially significant when $\tau_p$ is equal to the expected number of active devices and for longer sequence lengths. The simulation parameters used in the simulations are summarized in Table \ref{tbl:SysParameters}.
 
\begin{figure}[t!]
	\begin{center}
		\includegraphics[trim=.3cm 0cm 0cm 0.4cm,clip=true,width = 9cm]{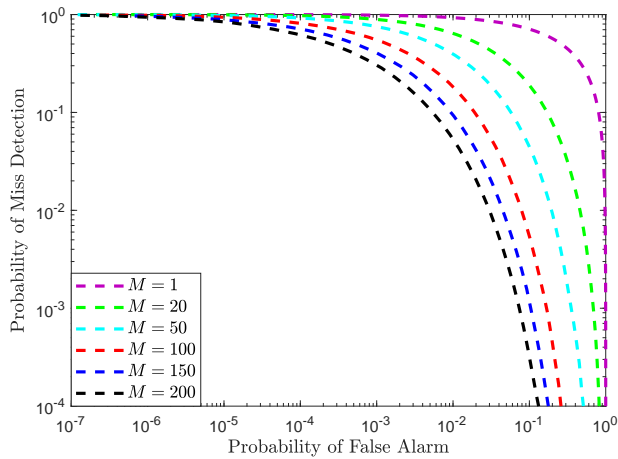}
		\caption{Probabilities of miss and false alarm for different numbers of antennas, $M$, 
		 for $N = 200$ devices with a device access probability $\epsilon = 0.05$ and a pilot length $\tau_p = 10$.}
		\label{fig:fig2}
	\end{center}
\end{figure}

Another crucial parameter which affects the user detection performance, is the number of antennas at the BS. The user detection performance of the AMP algorithm with respect to various number of BS antennas is illustrated in Fig.~\ref{fig:fig2}. Increasing the number of antennas significantly improves the performance. However, the performance gains due to increased numbers of antennas experience a saturation effect, i.e., the improvement gradually decreases as $M$ increases. This shows that increasing the number of antennas enhances the performance of the AMP algorithm for user detection; however the number of antennas should not be considered as an absolute substitute for pilot sequence length.
 
\begin{figure}[t!]
	\begin{center}
		\includegraphics[trim=.3cm 0cm 0cm 0.4cm,clip=true,width = 9cm]{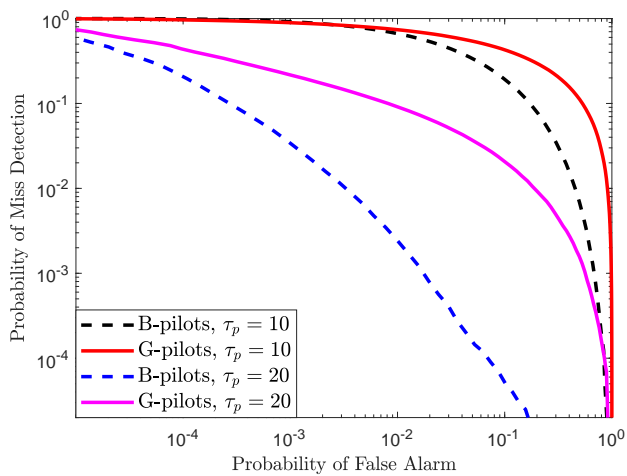}
		\caption{Probabilities of miss and false alarm using pilot sequences generated via Bernoulli (B-pilots) and Gaussian (G-pilots) distributions. The setup consists of  $M = 20$ BS antennas and $N = 200$ devices with a device access probability $\epsilon = 0.05$.}
		\label{fig:figNewPilots}
	\end{center}
\end{figure}
 
Pilot sequences generated by an i.i.d. complex Gaussian distribution represent another common choice for compressed sensing approaches \cite{liu2017massive}. Here, we have utilized pilot sequences generated by sampling an i.i.d. Bernoulli distribution. There are two reasons for this choice: First, it is easier and more practical to utilize sequences generated from a finite alphabet. Second, our numerical analysis demonstrates that Bernoulli sequences provides  better performance in terms of device activity detection. This is illustrated in Fig.~\ref{fig:figNewPilots}, which provides a comparison of device detection performance for different pilot sequences, using the AMP algorithm. The performance of Bernoulli sequences is better than that of Gaussian pilots, and the performance difference becomes more significant as the pilot length increases.

 \subsection{Asymptotic Analysis} \label{sec:Asymp}
 
The state evolution of the AMP algorithm is equivalent to applying a denoiser to a  signal received over an AWGN channel, in the asymptotic region. This property is shown to hold when the sensing matrix, $\bs{\Phi}$ in \eqref{eq:recCompMatrix}, is Gaussian. The state evolution is expected to hold for matrices with i.i.d. entries with zero mean and variance $1/\tau_p$. Even though there is numerical evidence that it holds for a broader class of matrices \cite{donoho2009message}, the characterization of the  matrices for which  the state evolution holds  is an open problem \cite{bayati2011dynamics}.   
 	
 In the rest of this section, we assume that state evolution holds in the asymptotic region, which allows us to provide a theoretical analysis of the device detection performance of AMP. Based on \eqref{eq:equaivalentAMP}, $\hat{\vect{x}}^t_n$ has i.i.d. Gaussian distributed elements with variance $\beta_n + \mu_t^2$, if $\alpha_n = 1$ and with variance $\mu_t^2$, if $\alpha_n = 0$. Here, $\mu_t^2$ denotes the diagonal elements of $\vect{\Sigma}^t$, which can be shown to be a diagonal matrix when the channels of a device across different antennas are assumed to be uncorrelated \cite{liu2017massive2}. Under these assumptions, we can state the following.

\begin{lemma}\label{lem:asymptoticLem}
	Assume that the detection of devices is carried out by comparing $\|\hat{\vect{x}}^t_n\|^2$ with a threshold, $\zeta$. Then, the miss detection and false alarm probabilities of the AMP algorithm, in the asymptotic region, for any threshold satisfying 
	\begin{equation}\label{eq:thrIntervalLem}
    M\mu_t^2 < \zeta < M\left(\beta_n + \mu_t^2\right)
	\end{equation}
	goes to zero when $ M\rightarrow \infty$:
	\begin{eqnarray}
	\lim\limits_{M\rightarrow \infty}\text{Pr}^{\text{MD}} \left(M, \zeta \right) &\rightarrow& 0, \\ 
	\lim\limits_{M\rightarrow \infty}\text{Pr}^{\text{FA}} \left(M, \zeta \right) &\rightarrow& 0.
	\end{eqnarray}     
\end{lemma}  
 \begin{IEEEproof}
 	See Appendix \ref{sec:AppAsym}.
 	\end{IEEEproof}

 Lemma \ref{lem:asymptoticLem}, states that perfect detection is possible in the asymptotic region. This is expected, since as $\tau_p \rightarrow \infty$ the pilot sequences become orthogonal which eliminates the cross-correlation between them. Moreover, as $M \rightarrow \infty$ the impact of noise also vanishes which allows for perfect detection. A similar analysis can be found in \cite{liu2017massive2}.
 
\begin{figure}[t!]
	\begin{center}
		\includegraphics[trim=.3cm 0cm 0cm 0.2cm,clip=true,width = 9cm]{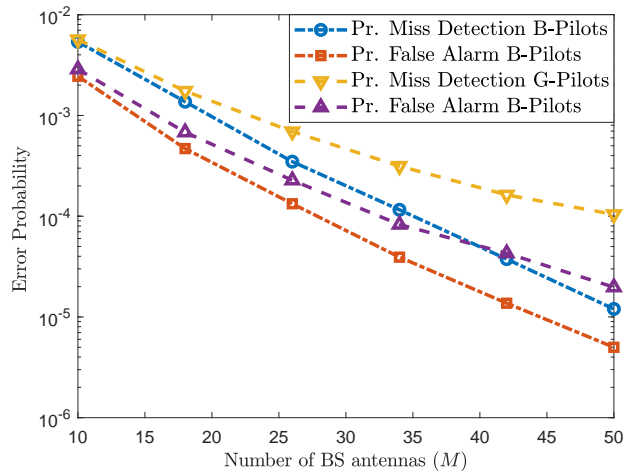}
		\caption{Error probabilities with respect to $M$, in  a setup with $N = 2000$ devices, device access probability of $\epsilon = 0.05$ and a pilot length of $\tau_p = 150$.}
		\label{fig:figAsympNew}
	\end{center}
\end{figure}

Fig.~\ref{fig:figAsympNew} illustrates the miss detection and false alarm probabilities as a function of $M$ for both Gaussian and Bernoulli sequences. Since the goal is to analyze the asymptotic behavior, we consider a setup with  a large number of devices,   $N = 2000$,   with a device access probability of $\epsilon = 0.05$ and pilot length  $\tau_p = 150$. In both cases, the detection performance improves with the number of BS antennas as predicted by Lemma \ref{lem:asymptoticLem}. An important point is that Bernoulli sequences provide better performance compared to Gaussian sequences. 
 
 \section{Power Control}  \label{sec:PowerControl}
 
 The assumption on identical transmission power, i.e., lack of power control, is common in compressed sensing approaches \cite{senel2017device,liu2017massive}, but this  is strictly suboptimal.
Simple power control strategies are  suitable for MTC scenarios with low-complexity, low-power devices. Especially, for the mMTC uplink, power control scenarios based on small-scale
 fading coefficients are not practical as accurate channel state information is difficult to acquire and inefficient for the transmission of small packages.  
 
 Gradually decreasing transmission power based on the large-scale fading, also referred to as   ``statistical channel inversion'' (SCI) \cite{massivemimobook}, 
 helps   reduce the channel gain differences between users and is especially beneficial to the users with relatively weaker channel gains. As a simple power control policy, we employ SCI and adjust the powers as follows: 
 \begin{flalign} \label{eq:SCI} 
 \rho_k = \rho_{ul}^{\max} \frac{\beta_{\min}}{\beta_k},
 \end{flalign}
 where $\rho_{ul}^{\max}$ is the maximum transmission power and $\beta_{\min}$ represents the minimum large-scale coefficient in the cell.  Using SCI, the device with the lowest large-scale coefficient will transmit at maximum power and the other devices' transmission powers scale inversely proportionally to  their large-scale coefficients.  Note that in practice there would have to be some signaling mechanism by which the base station informs the users about $\beta_{\min}$. If a user has a value of $\beta$ below $\beta_{\min}$, it would not be able to access the network using its available power budget.
 
 \begin{figure}[t!]
 	\begin{center}
 		\includegraphics[trim=.3cm 0cm 0cm 0.2cm,clip=true,width = 9cm]{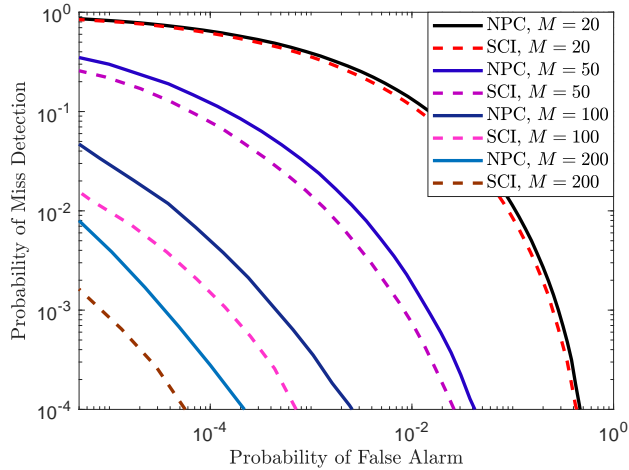}
 		\caption{Comparison of power control strategies, for different numbers of antennas, $M$, in  a setup with $N = 200$ devices, a device access probability of
		$\epsilon = 0.05$ and a pilot length of $\tau_p = 15$.}
 		\label{fig:fig6}
 	\end{center}
 \end{figure}
 
Fig.~\ref{fig:fig6} illustrates the performance difference between the two cases with no power control (NPC) and SCI. With NPC, each device transmits with maximum power, whereas with SCI each device adjusts its transmission power based on \eqref{eq:SCI}. An important difference between the two setups is that with SCI the total  power consumption is less. Hence, the total interference in the system is also higher with NPC than with  SCI. Fig.~\ref{fig:fig6} shows that even with a very simple power control policy, the device detection performance is improved. The difference is even more significant when the number of antennas is increased. In the subsequent numerical analysis, SCI is employed.

\section{Coherent Transmission} \label{sec:ChEstimation}

In the canonical massive MIMO setup with TDD operation, each coherence interval   consists of three phases: uplink training, uplink, and downlink data transmission. In this section, we focus on the uplink training and data transmission, and thus the downlink data transmission phase is neglected. The channel estimates acquired via uplink training are utilized at the BS during the uplink data transmission of devices.

The channel estimates provided by the AMP algorithm can be used for coherent data transmission along with device detection \cite{liu2017massive}. However, it is possible to obtain a more accurate channel estimate of device $k$ as follows,
\begin{eqnarray}
\vect{y}_k &=& \vect{Y}\bs{\varphi}_k = \sum_{k' \in \mathcal{K}}^{} \left(\sqrt{\rho_{k'}\tau_p}\vect{g}_{k'}\bs{\varphi}_{k'}^H + \vect{Z}\right)\bs{\varphi}_k, \nonumber \\
&=& \sqrt{\rho_{k}\tau_p}\vect{g}_k + \sum_{k'\in \mathcal{K} \setminus \{k\}}^{} \sqrt{\rho_{k'}\tau_p}\vect{g}_{k'}\bs{\varphi}_{k'}^H\bs{\varphi}_{k} + \vect{z}', \nonumber
\end{eqnarray}
where $\mathcal{K}$ is the set of active devices and $\vect{z}' = \vect{Z}\bs{\varphi}_k$ 
has i.i.d. $\mathit{CN} (0, \sigma^2)$ components as $\|\bs{ \varphi}_k\|^2 = 1$. Then, the LMMSE estimate of $\vect{g}_k$ is  
\begin{eqnarray}
\hat{\vect{g}}_k &=& \frac{\mathbb{E}\{\vect{y}^H_k\vect{g}_k\}}{\mathbb{E}\{\vect{y}_k^H\vect{y}_k\}}\vect{y}_k, \nonumber \\ 
&=&   \frac{\sqrt{\rho_{k}\tau_p}\beta_k}{\sum_{k'\in \mathcal{K}}^{}\rho_{k'}\tau_p\beta_{k'} |\bs{\varphi}_k^H\bs{\varphi}_{k'}|^2 + \sigma^2 } \vect{y}_k, \label{eq:estimateKnownDevices}
\end{eqnarray}
which only considers the effect of the active devices. Hence, 
once the set of active devices is determined (the non-zero rows of $\vect{X}$) by the AMP, 
the MMSE estimator can be utilized to obtain a channel estimate,  
which provides the true MMSE were the  pilot sequences and large-scale fading coefficients known at the BS.  
\begin{remark}
We assume perfect device detection in this section, since our focus is to demonstrate that it is possible to obtain more accurate channel estimates via MMSE, than what
the AMP algorithm delivers. Consequently,   higher rates are achievable by using the
 MMSE estimator after the device detection. The performance of coherent transmission without perfect device detection assumption is investigated later.
\end{remark}

 The complexity of the AMP is $O(NM\tau_p)$ per iteration. The increased complexity due to MMSE estimation is less than the equivalent of one iteration in the AMP algorithm. Note that 
 although  we explicitly considered the AMP  in this section, the ideas can be employed with any compressed sensing techniques, such as those in \cite{tropp2006algorithms}.

During the data transmission the BS receives,
\begin{align}
\vect{y} =  \sum_{k'\in \mathcal{K}}^{}\sqrt{\rho_{k'}}\vect{g}_{k'}x_{k'} + \vect{z} 
\end{align}  
where $x_{k}$ represents the data symbol of device $k$. Each device  transmits unit-power symbols, i.e., $\mathbb{E}\{|x_k|^2\} = 1$. To detect the data symbols of device $k$, the BS employs a combining vector, $\vect{v}_k$, as follows:
\begin{align}\label{eq:ReceivedDataSignalProcessed}
 \tilde{\vect{y}}_k= \vect{v}_k^H\vect{y} = \sum_{k'\in \mathcal{K}}^{}\sqrt{\rho_{k'}}\vect{v}_k^H\vect{g}_{k'}x_{k'} + \vect{v}_k^H\vect{z}.   
\end{align}
Based on \eqref{eq:ReceivedDataSignalProcessed}, an ergodic achievable rate of device $k$ is 
\begin{equation}
R_k = \log_2 (1 + \Gamma_k), 
\end{equation}
where
\begin{equation}\label{eq:SINReffective}
\Gamma_k = \frac{|\mathbb{E}\left(\vect{v}_k^H\vect{g}_k\right)|^2\rho_{k}}{\sum\limits_{k'\in \mathcal{K}}^{}{\mathbb{E}\left(|\vect{v}_k^H\vect{g}_{k'}|^2\right)}\rho_{k'} + \mathbb{E}\left(\|\vect{v}_k\|^2\right)\sigma^2 - |\mathbb{E}\left(\vect{v}_k^H\vect{g}_k\right)|^2\rho_{k}}.
\end{equation} 
Consider the MRC vector, 
\begin{align}\label{eq:MRCvector}
\vect{v}_k = \frac{1}{\gamma_{k}\sqrt{M}} \hat{\vect{g}}_k,
\end{align} 
where $\gamma_{k}$ is the mean square of the $m$-th element of $\hat{\vect{g}}_k$, given by  
\begin{align}\label{eq:MSgamma}
\gamma_{k} &= \mathbb{E}\left[\left|\left[\hat{\vect{g}}_k\right]_m\right|^2\right], \\ &= \frac{\rho_{k}\tau_p\beta_k^2}{\sum_{k'\in \mathcal{K}}^{}\beta_{k'}\rho_{k'}\tau_p |\bs{\varphi}_k^H\bs{\varphi}_{k'}|^2 + \sigma^2}.
\end{align} 
Then, the spectral efficiency with MRC can be computed using the bounding techniques given in \cite[Sec.~2.3.4]{redbook}, giving  the following result.
\begin{lemma}\label{lem:SE-1}
An achievable rate of device $k$ is given by, 
\begin{align}
R_k = (1 - \frac{\tau_p}{\tau}) \log_2\left(1 + \Gamma_k\right)
\end{align}	
where $\Gamma_k$ is the effective SINR given by
\begin{align} \label{eq:SINR-MRC}
\Gamma_k = \frac{M\rho_k}{M\hspace{-.4cm}\sum\limits_{k'\in \mathcal{K}\backslash\{k\}}^{}\hspace{-.3cm}\frac{|\bs{\varphi}_k^H\bs{\varphi}_{k'}|^2\rho_{k'}^2\beta_{k'}^2}{\rho_k\beta_k^2} + \frac{1}{\gamma_{k}}\left(\sum\limits_{k'\in \mathcal{K}}^{} \rho_{k'}\beta_{k'} + \sigma^2\right)  }.
\end{align}

\end{lemma}

\begin{IEEEproof}
See Appendix \ref{sec:AppRate}.
\end{IEEEproof}	

\begin{remark}
	Although there are other methods (e.g., zero-forcing and MMSE), we only consider MRC throughout this work for conciseness, and  as the performance of different combining techniques is not in our focus. Detailed formulas for   performance of other detection techniques follow
	by direct application of techniques in
	 \cite{redbook}. 
\end{remark}

The rate that can be achieved with coherent transmission is limited by the non-orthogonality of the pilots, which creates
coherent interference that scales with the number of antennas. Especially, in the asymptotic region when $M \rightarrow \infty$, the effective SINR defined by \eqref{eq:SINR-MRC} becomes
\begin{align} \label{eq:SINR-MRCasymp}
\Gamma_k = \frac{\rho_k^2\beta_k^2}{\hspace{-.4cm}\sum\limits_{k'\in \mathcal{K}\backslash\{k\}}^{}\hspace{-.3cm}{|\bs{\varphi}_k^H\bs{\varphi}_{k'}|^2\rho_{k'}^2\beta_{k'}^2} }.
\end{align}
In this regime, the non-orthogonality of the pilots is the limiting factor for the achievable rate. 

 An important point is that \eqref{eq:SINR-MRC} is valid for long block lengths while for control signaling tasks, probability of error is a more relevant performance measure.  Nevertheless,
the ergodic capacity gives an indication of how, qualitatively at least, the performance varies with the different system parameters. Other performance metrics such as maximum coding rate, are available for short packet lengths in the literature on finite-block length information theory \cite{polyanskiy2010channel}. However, for control signaling applications where only a few bits are to be transmitted, the finite-block length bounds are not tight and 
error probability is a more reliable performance metric.

\begin{figure}[t!]
	\begin{center}
		\includegraphics[trim=.3cm 0cm 0cm 0.2cm,clip=true,width = 9cm]{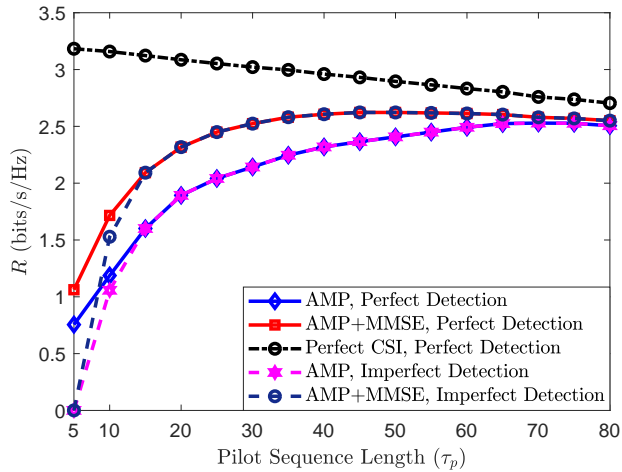}
		\caption{Comparison of achievable ergodic
		rates with coherent detection, with  different channel estimators and with MRC for  a setup with $M = 50$ antennas, $N = 100$ devices, a device access
		probability of $\epsilon = 0.05$, and a coherence interval length of $\tau = 500$ symbols.}
		\label{fig:fig5}
	\end{center}
\end{figure}

Fig.~\ref{fig:fig5} illustrates how the quality of the channel estimates impacts the spectral efficiency. The rates shown take the pilot overhead into account, and hence represent ``net throughputs'' (per unit bandwidth and time unit). Specifically, the total number of symbols available for pilots and data transmission is fixed, which results in fewer data symbols as the
pilot sequence length increases.
The estimates obtained via AMP and MMSE (after device detection is accomplished by AMP) both with and without perfect device detection assumption are compared with the perfect CSI case. In the perfect CSI case, the active devices are also assumed to be perfectly detected at the BS. The difference between the cases with and without the perfect device detection assumption, vanishes quickly with increasing pilot sequence length and around $\tau_p = 20$, the difference becomes negligible. Moreover, the difference between AMP and MMSE estimates also vanishes as the pilot sequence length is increased. Similarly both techniques approach the perfect CSI case, since in the asymptotic region ($\tau_p \rightarrow \infty$), the effect of noise on the channel estimates vanishes and the pilot sequences become orthogonal. In the perfect CSI case, the rate decreases with the pilot sequence length as the pre-log term decreases with $\tau_p$. Also note that the number of data symbols available for both AMP and AMP+MMSE is identical, as MMSE estimation is carried out based on the non-orthogonal pilots used for device detection.

\section{Non-coherent Transmission} \label{sec:non-coh}


In contrast to coherent transmission, explicit channel estimates are not required with non-coherent transmission. In order to convey $r$ bits of data non-coherently, each device is allocated $2^r$ distinct pilot sequences and transmits one of these sequences based on the $r$ information bits. Here, the information is embedded into the pilot sequences and there is no need to allocate additional symbols for data transmission. Hence, all $\tau$ symbols can be utilized for pilot sequences.

Let $\bs{\bar{\Phi}}_k = [\bs{\varphi}_{k,0}, \bs{\varphi}_{k,1}, \ldots, \bs{\varphi}_{k,2^r -1}] \in \mathbb{C}^{\tau  \times 2^r} $  denote the pilot sequences allocated for device $k$. This device transmits exactly only one of these pilot sequences, selected based on the information bits. Then, the composite received signal at the BS is       
\begin{equation} \label{eq:recCompMatrixEIB}
\vect{Y} = \sqrt{\rho_{ul} \tau_p} \bs{\bar{\Phi}}\bar{\vect{X}} + \vect{Z},
\end{equation}
where $$\bs{\bar{\Phi}} = [\bs{\bar{\Phi}}_1, \ldots, \bs{\bar{\Phi}}_N] \in \mathbb{C}^{\tau  \times N2^r}$$ and $$\bar{\vect{X}} = [\bar{\vect{X}}_1, \ldots, \bar{\vect{X}}_N]^H$$ where $$\bar{\vect{X}}_k = [\alpha_{k,1}\vect{g}_k, \ldots, \alpha_{k,2^r}\vect{g}_k] \in \mathbb{C}^{M\times 2^r}.$$ Here $\alpha_{k,l} = 1$ if device $k$ is active and the $l$th-symbol is  
embedded. Recall that each device is active with probability $\epsilon$ and
\begin{flalign}
\sum_{i = 0}^{2^r -1} \alpha_{n,i} = \begin{cases}1, & \text{with Pr.}~\epsilon,\\ 0, & \text{with Pr.}~ 1- \epsilon, 
\end{cases}~~~\forall n = 1, \ldots, N.
\end{flalign}
Notice that with non-coherent transmission, the BS must consider $N 2^r$ pilot sequences instead of $N$.
However, the number of active users remains the same, i.e., the number of non-zero rows of $\bar{\vect{X}}$ and $\vect{X}$ is equal. The active devices along with their embedded bits could in principle be detected by the AMP algorithm without any modification, implicitly   assuming each pilot sequence is associated with  a different, fictitious device. But such an approach is strictly sub-optimal, as the available information about the structure of $\bar{\vect{X}}$ is not utilized. A modified AMP algorithm for the case of $r=1$ was outlined in \cite{senel2017device}. Here, we present its extension to the general $r$-bit case.  

\subsection{Algorithm Description} \label{sec:AlgDes}

Assigning multiple pilot sequences to a device increases the sparsity, i.e., the number of non-zero rows of $\bar{\vect{X}}$ and $\vect{X}$ are equal; however $\bar{\vect{X}}$ has $2^r$ times more rows than $\vect{X}$. This increase in the sparsity manifests itself structurally in $\bar{\vect{X}}$, as it is impossible to have multiple non-zero rows corresponding to the same device. In order to exploit these new structural properties of $\tilde{\vect{X}}$, we propose a modified AMP algorithm to be used for the detection of embedded bits along with device detection. 

Let $\bar{\vect{X}}_k = [\bar{\vect{x}}_{k,1}, \ldots, \bar{\vect{x}}_{k,2^r}] \in \mathbb{C}^{M\times 2^r}$ and $\hat{\bar{\vect{x}}}_{k,l}$ be the estimate of the row of $\bar{\vect{X}}$ corresponding to the $l$th pilot sequence of device $k$. Assume that user $k$ is active and transmitting the pilot sequence $l'$, i.e., $\alpha_{k, l'} = 1$; then 
\begin{flalign} \label{eq:2casesX} 
\hat{\bar{\vect{x}}}^t_{k, l} = \begin{cases} 
\vect{g}_k + (\vect{\Sigma}^t)^{\frac{1}{2}}\vect{w}\sim \mathit{CN}(0, \beta_k \vect{I}+ \vect{\Sigma}^t), & \text{if}~~ l = l', \\
(\vect{\Sigma}^t)^{\frac{1}{2}}\vect{w}\sim \mathit{CN}(0, \vect{\Sigma}^t), & \text{if}~~ l \neq l'.
\end{cases}
\end{flalign}
Hence only a single row corresponding to device $k$ is non-zero. The likelihood function based on \eqref{eq:2casesX} is given by
\begin{equation}
\Lambda (\hat{\bar{\vect{x}}}^t_{k,l}) = \frac{|\vect{\Sigma}^t|}{|\beta_k \vect{I}+\vect{\Sigma}^t|} q(\hat{\bar{\vect{x}}}^t_{k,l}; \vect{\Sigma}^t )^{-1}.
\end{equation}
Let $\varphi(\hat{\bar{\vect{x}}}^t_{k,l})$ denote the sequence likelihood fraction (SLF) coefficient defined by 
\begin{equation}
\varphi(\hat{\bar{\vect{x}}}^t_{k,l}) = \frac{\Lambda (\hat{\bar{\vect{x}}}^t_{k,l})}{\sum_{l' = 1}^{2^r}\Lambda (\hat{\bar{\vect{x}}}^t_{k,l'}) }.
\end{equation}
This coefficient  can be thought of as a measure of the proportional likelihood of a given sequence allocated to device $k$. The SLF coefficient provides a form of proportional thresholding; however in order to enhance its effectiveness, a sharper threshold is required. In the ideal case, the receiver should only decide on one of the possible pilot sequences while suppressing the other one. In order to achieve this, we utilize a soft-thresholding function known as a sigmoid function. More specifically, the sigmoid function is defined by 
\begin{equation}\label{eq:sigmoid}
f(x) = \frac{1}{1 + \exp(-c(x - \frac{1}{2}))}.
\end{equation}   
where $c$ is a parameter that determines the  sharpness of the sigmoidal transition. 
  The resulting modified denoiser is 
\begin{equation}\label{eq:modifiedEta}
\tilde{\eta}_{t,n}(\hat{\bar{\vect{x}}}_n^t) = f(\varphi(\hat{\bar{\vect{x}}}^t_k))\eta_{t,n}(\hat{\bar{\vect{x}}}_n^t).
\end{equation}

Note that the modified denoiser is Lipschitz-continuous. However, the validity of state evolution is unclear, as the unmodified case with Bernoulli sequences is only verified via numerical analysis. Even though there are some results for the cases where the modifying function is separable and the sensing matrix has a special structure \cite{bayati2015universality}, the asymptotic behavior of AMP with other   sensing matrix distributions than Gaussian, is an open problem.     

The proposed modified AMP algorithm 
(M-AMP) is specifically designed for non-coherent transmission. The principal idea is that only a single row corresponding to a device may be non-zero as it is impossible for a device to transmit both pilot sequences concurrently.

\begin{figure}[thb]
	\begin{center}
		\includegraphics[trim=0.5cm 0cm 0cm 0.3cm,clip=true,width = 9.5cm]{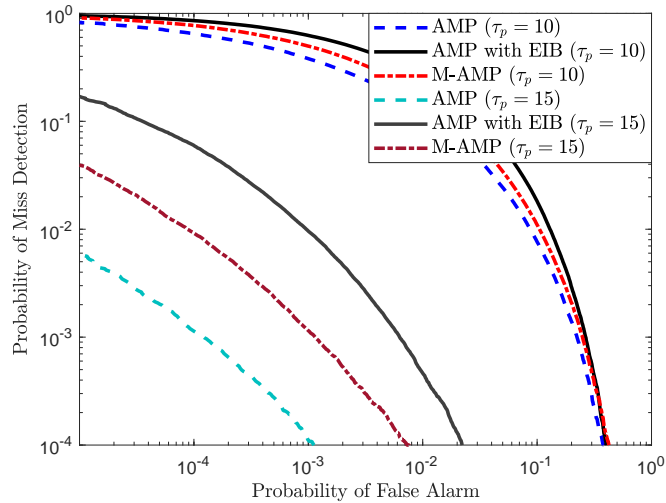}
		\caption{Probabilities of miss and false alarm of device activity detection, for various pilot lengths, $\tau_p$,  in a setup with $N = 100$ devices, $M = 50$ antennas and a device
		access probability of $\epsilon = 0.1$.}
		\label{fig:fig8}
	\end{center}
\end{figure}

In Fig. \ref{fig:fig8}, the user detection performances of three different AMP algorithms are depicted. The algorithms compared are as follows:
\begin{itemize}
	\item \textbf{AMP}: The original AMP algorithm which considers $N = 100$ pilot sequences without any embedded information bit.
	
	\item \textbf{AMP with EIB}: The original AMP algorithm which considers $N =200$ pilot sequences and detects users along with a single embedded information bit. 
	
	\item \textbf{M-AMP}: The modified AMP algorithm which considers $N=200$ pilot sequences and detects users along with a single embedded information bit.
\end{itemize}
There are $100$ potential users and on   average only $\epsilon N$ are active. For the case when a single information bit is transmitted, the detector must consider $200$ pilot sequences. In this case, if the detector determines that one of the pilot sequences corresponding to a user is transmitted, then that user is detected as an active user independently of whether an information bit is transmitted. In all cases, the number of iterations and pilot sequence length are identical. As expected the AMP algorithm without any additional information bit provides the best performance.   M-AMP outperforms the original AMP when the embedded information bit is to be detected along with the device activity. The performance difference between the algorithms becomes more significant with increased pilot length.  
 
\begin{figure}[t!]
	\begin{center}
		\includegraphics[trim=0.5cm 0cm 0cm 0.3cm,clip=true,width = 9.5cm]{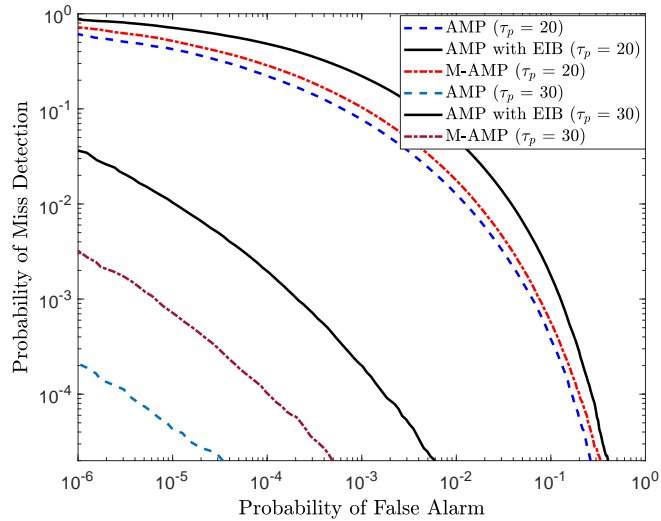}
		\caption{Probabilities of miss and false alarm of device activity detection, for various
		pilot lengths, $\tau_p$, for a setup with $N = 200$ devices, $M = 50$ antennas
		and a device activity probability of $\epsilon = 0.1$.\newline
		\label{fig:fig9}}
	\end{center}
\end{figure}
 
An interesting property of the AMP algorithm is that increasing $N$, $\tau_p$ and $K$ while keeping their ratios fixed improves the performance. Fig. \ref{fig:fig9} illustrates the scaling of the device detection performance of the three approaches. The behavior of each algorithm is similar; however the performances of all of the approaches are superior compared to the case with $100$ users. This is a desirable property in  mMTC scenarios with large numbers of devices.

\subsection{Coherent versus Non-Coherent Transmission} \label{sec:CohVsNon}

In this section, we compare  coherent and non-coherent transmission for an mMTC scenario where each device aims to transmit a few data bits. No prior information on the set of active devices is assumed. Since, the goal is to convey a small number of bits of data, we utilize probability of error as a performance metric. 

\begin{figure}[t!]
	\begin{center}
		\includegraphics[trim=.5cm 0cm 0.1cm 0.6cm,clip=true, width = 9.5cm]{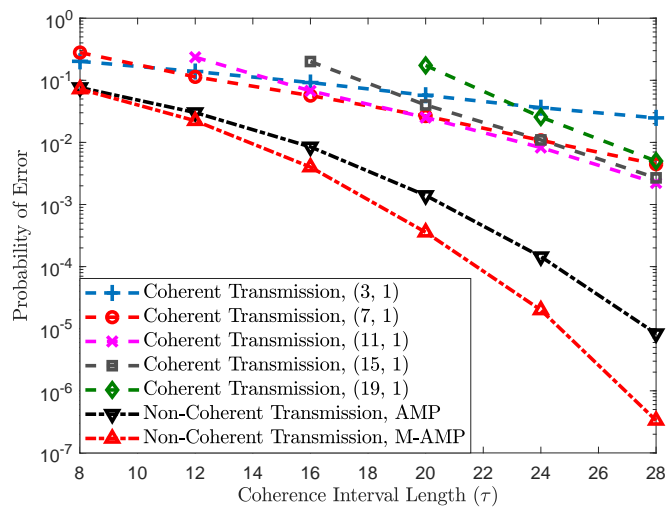}
		\caption{Probability of error for the transmission of a single embedded information bit,
		as function of
		 the coherence interval length, $\tau$, for various (repetition) code lengths in
		  a setup with 
			$M = 20$ antennas,  $N = 100$ devices, and a device activity probability
			of $\epsilon = 0.1$.}
		\label{fig:fig12}
	\end{center}
\end{figure}

\begin{figure}[t!]
	\begin{center}
		\includegraphics[trim=.6cm 0cm 0.2cm 0.1cm,clip=true,width = 9.5cm]{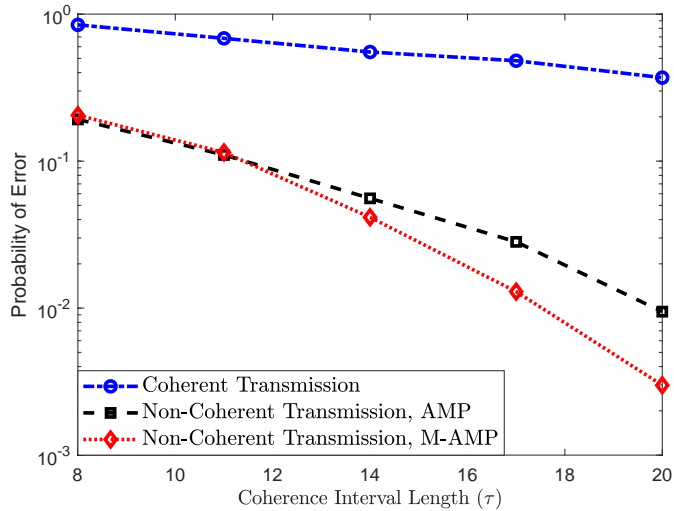}
		\caption{Probability of error for the transmission of $4$
		embedded information bits in a setup with $M = 20$ antennas, $N = 100$ 
		devices and a device activity probability $\epsilon = 0.1$. }
		\label{fig:fig11}
	\end{center}
\end{figure}

Fig.~\ref{fig:fig12} illustrates the performance of coherent and non-coherent transmission in terms of probability of error for a single bit of information, for different  coherence interval lengths. For coherent transmission, first AMP is employed to detect the active devices and obtain the channel estimates. Then, a repetition code of varying rate with BPSK transmission is employed to convey a single bit of information. Hence, entire coherence interval except for the repetition-coded information
bit  is utilized as pilot sequence. The best performance is obtained with a length-$11$ repetition code, whereas lengths $15$ and $19$ provide similar performances. 
For the non-coherent transmission, AMP and M-AMP are employed to detect the transmitted pilot sequences among $2N = 200$ candidates. The results shows that non-coherent transmission
not only outperforms coherent transmission but also scales better with the coherence interval length.

Fig.~\ref{fig:fig11} illustrates the performance of coherent and non-coherent transmission in terms of probability of error for transmission of $4$-bits with respect to coherence interval length. For coherent transmission, a $(7,4)$-Hamming code is utilized to convey $4$ information bits after the channel estimates are acquired. The original and the modified AMP algorithms are utilized to detect the active users and the transmitted pilot sequences among $16N = 1600$ candidates. Even though non-coherent transmission provides  significantly better performance, as the number of information bits increases, this difference vanishes. The performance difference between coherent and non-coherent transmission is more significant for the single bit case shown in Fig.~\ref{fig:fig12} compared to Fig.~\ref{fig:fig11}.

\section{Conclusion}

We investigated the joint device detection and data transmission problem in an mMTC setup,
where devices use non-orthogonal pilots. The device detection is carried out using the AMP  compressed sensing algorithm. A simple power control technique which only relies on large-scale coefficients is employed   and  shown to enhance the performance. 
We also showed that  once the active devices have been detected, it is possible to obtain more accurate channel estimates by using  MMSE estimation, instead of relying on the estimates provided by the AMP. This in turn results in a higher spectral efficiency for coherent transmission. 

Targeting the vision of fully non-coherent communication for mMTC in Massive MIMO, especially for control signaling, we furthermore proposed a novel non-coherent transmission scheme. This scheme   encodes the information to be transmitted into the choice of pilot sequence sent by each devices, specifically mapping $r$ information bits onto $2^r$ possible pilots per device.
We devised a modified AMP (M-AMP) algorithm designed specifically to exploit the structured sparsity incurred by the proposed non-coherent transmission scheme. The M-AMP 
algorithm not only outperforms the original AMP algorithm for  the non-coherent scheme, but also scales better with the number of devices. 
A comparison of coherent and non-coherent transmission revealed that non-coherent transmission significantly outperforms the coherent transmission scheme. This suggests that the  proposed non-coherent transmission approach  can be useful  in future mMTC networks.


%
 
\ifCLASSOPTIONcaptionsoff
\newpage
\fi

\appendices
\section{Proof of Lemma \ref{lem:asymptoticLem}} \label{sec:AppAsym}
First, note that $\hat{\vect{x}}^t_n$ defined by \eqref{eq:equaivalentAMP}, is a random vector where each element has i.i.d. Gaussian distributed real and imaginary parts. Hence,  $\|\hat{\vect{x}}^t_n\|^2/((\beta_n + \mu_t^2))$ given $\alpha_n = 1$ and  $\|\hat{\vect{x}}^t_n\|^2/ \mu_t^2$ given $\alpha_n = 0$ follows a $\chi^2$ distribution with 2$M$ degrees of freedom (DoF). The cumulative distribution function is defined by 
\begin{equation}\label{eq:gammaCDF}
\text{Pr}\left(\|\hat{\vect{x}}^t_n\|^2 \leq \zeta \right) = \frac{\gamma(M,\zeta/2)}{\Gamma(M)},
\end{equation}
where $\Gamma(\cdot)$ represents the Gamma function and $\gamma(\cdot)$ is the lower incomplete Gamma function. Since, our focus is on asymptotic behavior, i.e., $\tau_p,~K,~N \rightarrow \infty$ while their ratios are fixed, the probabilities of miss detection and false alarm as a function of number of BS antennas, $M$, can be defined as follows   
\begin{eqnarray}
\text{Pr}^{\text{MD}} \left(M, \zeta \right) &=& \text{Pr}\left(\|\hat{\vect{x}}^t_n\|^2 \leq \zeta_{\text{MD}} | \alpha_n = 1\right) = \frac{\gamma(M,\zeta_{\text{MD}}/2)}{\Gamma(M)}, \label{eq:probMiss} \\
\text{Pr}^{\text{FA}} \left(M, \zeta \right) &=& \text{Pr}\left(\|\hat{\vect{x}}^t_n\|^2 > \zeta_{\text{FA}} | \alpha_n = 0\right) = 1 - \frac{\gamma(M,\zeta_{\text{FA}}/2)}{\Gamma(M)}. \label{eq:probFalseAlarm}
\end{eqnarray}  
where 
\begin{eqnarray}
\zeta_{\text{MD}} &=& \frac{\zeta}{\beta_n + \mu_t^2}, \label{eq:thrMD}\\
\zeta_{\text{FA}} &=& \frac{\zeta}{\mu_t^2} \label{eq:thrFA}.
\end{eqnarray}
 
 Equations \eqref{eq:probMiss} and \eqref{eq:probFalseAlarm} define the probabilities of miss detection and false alarm in terms of Gamma functions. First, we focus on the probability of miss detection and use an asymptotic representation of the type  
\begin{equation}\label{eq:lowGammExp}
   \frac{\gamma(M,\zeta_{\text{MD}}/2) }{\Gamma(M)} = \frac{1}{2}\erfc\left(-\underline{\eta}\sqrt{M/2}\right) - R_M(\underline{\eta}),  
\end{equation}
where, 
\begin{equation}\label{eq:Rm}
R_M(\underline{\eta}) \sim \frac{\exp\left(-\frac{1}{2} M\underline{\eta}^2  \right) }{\sqrt{2\pi M}} \sum_{i = 0}^{\infty}\frac{c_i(\underline{\eta})}{M^i}, \quad M\rightarrow \infty,
\end{equation}
which is derived in \cite{gautschi1998incomplete}. Here, $\underline{\eta} = \sqrt{2(\lambda_{\text{MD}} - 1 - \ln\lambda_{\text{MD}})}$ for $\lambda_{\text{MD}} = \zeta_{\text{MD}}/M < 1$ and the first coefficient $c_0(\underline{\eta})$ is defined by 
\begin{equation}\label{eq:coeffApp}
c_0(\underline{\eta}) = \frac{1}{\lambda_{\text{MD}} -1} - \frac{1}{\underline{\eta}}, 
\end{equation}
and the remaining terms of $R_M$ are at least of $o(\exp(-M)/M^{3/2})$ \cite{temme1979asymptotic}. Finally, we use the following approximation for the $\erfc(\cdot)$ function
\begin{equation}\label{eq:erfcApp}
\erfc(x) = \frac{\exp(-x^2)}{\sqrt{\pi}x}\left(1 + o\left(\frac{1}{x^2}\right)\right).
\end{equation}
Hence, \eqref{eq:lowGammExp} can be re-written as
\begin{eqnarray}
\text{Pr}^{\text{MD}} \left(M, \zeta_{\text{MD}}/2 \right) &=& \frac{1}{2}\frac{\exp\left( -\underline{\eta}^2M/2\right)}{-\underline{\eta}\sqrt{\pi M/2}} \left(1 + o\left(\frac{1}{M}\right)\right) - \frac{\exp\left(-\underline{\eta}^2M/2\right)}{\sqrt{2\pi M}} \left(\frac{1}{\lambda_{\text{MD}} - 1 } - \frac{1}{\underline{\eta}}\right)\nonumber \\ &-& o\left(\frac{\exp\left(-M\right)}{M\sqrt{M}}\right), \nonumber \\ 
&=& \frac{\exp\left(-\underline{\eta}^2M/2\right)}{\sqrt{2\pi M}}\frac{1}{1-\lambda_{\text{MD}}} + o\left(\frac{1}{M}\right) = o\left(\frac{1}{M}\right). \label{eq:PrMissTh}
\end{eqnarray}

Similarly, for the false alarm detection, we start with the asymptotic representation of type
\begin{equation}\label{eq:uppGammExp}
1 - \frac{\gamma(M,\zeta_{\text{FA}}/2) }{\Gamma(M)} = \frac{1}{2}\erfc\left(\overline{\eta}\sqrt{M/2}\right) + R_M(\overline{\eta}),  
\end{equation}
where $R_M(\overline{\eta})$ is defined by \eqref{eq:Rm} and $\overline{\eta} = -\sqrt{2(\lambda_{\text{FA}} - 1 - \ln\lambda_{\text{FA}})}$ for $\lambda_{\text{FA}} = \zeta_{\text{FA}}/M > 1$. Using \eqref{eq:Rm} and \eqref{eq:erfcApp} in \eqref{eq:uppGammExp}, we obtain
\begin{eqnarray}
\text{Pr}^{\text{FA}} \left(M, \zeta_{\text{FA}}/2 \right) &=& \frac{1}{2}\frac{\exp\left( -\overline{\eta}^2M/2\right)}{\overline{\eta}\sqrt{\pi M/2}} \left(1 + o\left(\frac{1}{M}\right)\right) + \frac{\exp\left(-\overline{\eta}^2M/2\right)}{\sqrt{2\pi M}} \left(\frac{1}{\lambda_{\text{FA}} - 1 } - \frac{1}{\overline{\eta}}\right) \nonumber \\ &+& o\left(\frac{\exp\left(-M\right)}{M\sqrt{M}}\right), \nonumber \\ 
&=& \frac{\exp\left(-\overline{\eta}^2M/2\right)}{\sqrt{2\pi M}}\frac{1}{\lambda_{\text{FA}}-1} + o\left(\frac{1}{M}\right) = o\left(\frac{1}{M}\right).
\end{eqnarray}
Hence, as  $M \rightarrow \infty$, $\text{Pr}^{\text{MD}} \left(M, \zeta \right)\rightarrow 0$, $\text{Pr}^{\text{FA}} \left(M, \zeta \right)  \rightarrow 0$ for any choice of 
\begin{equation}\label{eq:thrCondition}
\frac{\zeta_{\text{MD}}}{M} < 1 < \frac{\zeta_{\text{FA}}}{M}.
\end{equation}
Using \eqref{eq:thrMD} and \eqref{eq:thrFA}, the set of thresholds satisfying \eqref{eq:thrCondition} lies in the interval, 
\begin{equation}\label{eq:thrInterval}
M\mu_t^2 < \zeta < M\left(\beta_n + \mu_t^2\right).
\end{equation}
Hence, any choice of threshold satisfying \eqref{eq:thrInterval} will result in perfect detection in the asymptotic region which concludes the proof.

\section{Proof of Lemma \ref{lem:SE-1} }\label{sec:AppRate} 
The terms in \eqref{eq:SINReffective} can be computed as follows
\begin{eqnarray}
\left|\mathbb{E}\left(\vect{v}_k^H\vect{g}_k\right)\right|^2\rho_{k} &=& \left|\mathbb{E}\left(\frac{1}{\gamma_{k}\sqrt{M}}\hat{\vect{g}}_k^H\vect{g}_k\right)\right|^2\rho_{k} \nonumber \\
&=& M\rho_{k} \label{eq:lem1-pr1}
\end{eqnarray}
and 
\begin{eqnarray}
\sum\limits_{k'\in \mathcal{K}}^{}{\mathbb{E}\left(|\vect{v}_k^H\vect{g}_{k'}|^2\right)}\rho_{k'} = \sum\limits_{k'\in \mathcal{K}}^{}{\mathbb{E}\left(\left|\frac{1}{\gamma_{k}\sqrt{M}}\hat{\vect{g}}_k^H\vect{g}_{k'}\right|^2\right)}\rho_{k'}  \nonumber \\
= \sum\limits_{k'\in \mathcal{K}}^{}\frac{\beta_{k'}}{\gamma_{k}} + \frac{M\beta_{k'}^2\rho_{k'}^2 |\bs{\varphi}_k^H\bs{\varphi}_{k'}|^2}{\rho_{k}\beta_k^2}. \label{eq:lem1-pr2}
\end{eqnarray}
Finally, the noise term is 
\begin{equation}
\mathbb{E}\left(\|\vect{v}_k\|^2\right)\sigma^2 = \frac{\sigma^2}{\gamma_{k}}, \label{eq:lem1-pr3}
\end{equation}
and \eqref{eq:SINR-MRC} is obtained by simply substituting \eqref{eq:lem1-pr1}, \eqref{eq:lem1-pr2}, \eqref{eq:lem1-pr3} into \eqref{eq:SINReffective}.


\begin{thebibliography}{10}
	\providecommand{\url}[1]{#1}
	\csname url@samestyle\endcsname
	\providecommand{\newblock}{\relax}
	\providecommand{\bibinfo}[2]{#2}
	\providecommand{\BIBentrySTDinterwordspacing}{\spaceskip=0pt\relax}
	\providecommand{\BIBentryALTinterwordstretchfactor}{4}
	\providecommand{\BIBentryALTinterwordspacing}{\spaceskip=\fontdimen2\font plus
		\BIBentryALTinterwordstretchfactor\fontdimen3\font minus
		\fontdimen4\font\relax}
	\providecommand{\BIBforeignlanguage}[2]{{%
			\expandafter\ifx\csname l@#1\endcsname\relax
			\typeout{** WARNING: IEEEtran.bst: No hyphenation pattern has been}%
			\typeout{** loaded for the language `#1'. Using the pattern for}%
			\typeout{** the default language instead.}%
			\else
			\language=\csname l@#1\endcsname
			\fi
			#2}}
	\providecommand{\BIBdecl}{\relax}
	\BIBdecl
	\renewcommand{\BIBentryALTinterwordstretchfactor}{4}
	
	\bibitem{senel2017device}
	K.~Senel and E.~G. Larsson, ``{Device Activity and Embedded Information Bit
		Detection Using AMP in Massive MIMO},'' in \emph{IEEE Global Communications
		Conference (GLOBECOM)}, 2017.
	
	\bibitem{senel2017mMTCwsa}
	------, ``{Joint User Activity and Non-Coherent Data Detection in mMTC-Enabled
		Massive MIMO Using Machine Learning Algorithms},'' in \emph{International
		Workshop on Smart Antennas (WSA)}, 2018.
	
	\bibitem{deliverable2015d6}
	P.~Popovski \emph{et~al.}, ``{Final report on the METIS 5G system concept and
		technology roadmap},'' \emph{METIS Document ICT-317669-METIS/D6.6}, 2015.
	
	\bibitem{ericsson}
	\BIBentryALTinterwordspacing
	Ericsson, ``{5G systems: Enabling the transformation of industry and
		society},'' White Paper. [Online]. Available:
	\url{www.ericsson.com/assets/local/publications/white-papers/wp-5g-systems.pdf}
	\BIBentrySTDinterwordspacing
	
	\bibitem{centenaro2016long}
	M.~Centenaro, L.~Vangelista, A.~Zanella, and M.~Zorzi, ``{Long-range
		communications in unlicensed bands: The rising stars in the IoT and smart
		city scenarios},'' \emph{IEEE Wireless Communications}, vol.~23, no.~5, pp.
	60--67, 2016.
	
	\bibitem{bockelmann2016massive}
	C.~Bockelmann, N.~Pratas, H.~Nikopour, K.~Au, T.~Svensson, C.~Stefanovic,
	P.~Popovski, and A.~Dekorsy, ``{Massive machine-type communications in 5G:
		Physical and MAC-layer solutions},'' \emph{IEEE Communications Magazine},
	vol.~54, no.~9, pp. 59--65, 2016.
	
	\bibitem{boccardi2014five}
	F.~Boccardi, R.~W. Heath, A.~Lozano, T.~L. Marzetta, and P.~Popovski, ``{Five
		disruptive technology directions for 5G},'' \emph{IEEE Communications
		Magazine}, vol.~52, no.~2, pp. 74--80, 2014.
	
	\bibitem{larsson2012piggybacking}
	E.~G. Larsson and R.~Moosavi, ``Piggybacking an additional lonely bit on
	linearly coded payload data,'' \emph{IEEE Wireless Communications Letters},
	vol.~1, no.~4, pp. 292--295, 2012.
	
	\bibitem{dawy2017toward}
	Z.~Dawy, W.~Saad, A.~Ghosh, J.~G. Andrews, and E.~Yaacoub, ``Toward massive
	machine type cellular communications,'' \emph{IEEE Wireless Communications},
	vol.~24, no.~1, pp. 120--128, 2017.
	
	\bibitem{redbook}
	T.~L. Marzetta, E.~G. Larsson, H.~Yang, and H.~Q. Ngo, \emph{{Fundamentals of
			Massive MIMO}}.\hskip 1em plus 0.5em minus 0.4em\relax Cambridge University
	Press, 2016.
	
	\bibitem{sesia2011lte}
	S.~Sesia, M.~Baker, and I.~Toufik, \emph{{LTE-the UMTS long term evolution:
			from theory to practice}}.\hskip 1em plus 0.5em minus 0.4em\relax John Wiley
	\& Sons, 2011.
	
	\bibitem{de2017randomM}
	E.~De~Carvalho, E.~Bjornson, J.~H. Sorensen, P.~Popovski, and E.~G. Larsson,
	``{Random access protocols for massive MIMO},'' \emph{IEEE Communications
		Magazine}, vol.~55, no.~5, pp. 216--222, 2017.
	
	\bibitem{boljanovic2017user}
	\BIBentryALTinterwordspacing
	V.~Boljanovic, D.~Vukobratovic, P.~Popovski, and C.~Stefanovic, ``{User
		Activity Detection in Massive Random Access: Compressed Sensing vs. Coded
		Slotted ALOHA},'' 2017. [Online]. Available:
	\url{https://arxiv.org/abs/1706.09918}
	\BIBentrySTDinterwordspacing
	
	\bibitem{choi2017compressed}
	J.~W. Choi, B.~Shim, Y.~Ding, B.~Rao, and D.~I. Kim, ``{Compressed sensing for
		wireless communications: Useful tips and tricks},'' \emph{IEEE Commun.
		Surveys Tuts}, vol.~19, no.~3, pp. 1527--1549, 2017.
	
	\bibitem{nan2015efficient}
	Y.~Nan, L.~Zhang, and X.~Sun, ``{Efficient downlink channel estimation scheme
		based on block-structured compressive sensing for TDD massive MU-MIMO
		systems},'' \emph{IEEE Wireless Communications Letters}, vol.~4, no.~4, pp.
	345--348, 2015.
	
	\bibitem{choi2015downlink}
	J.~W. Choi, B.~Shim, and S.-H. Chang, ``{Downlink pilot reduction for massive
		MIMO systems via compressed sensing},'' \emph{IEEE Communications Letters},
	vol.~19, no.~11, pp. 1889--1892, 2015.
	
	\bibitem{du2017efficient}
	Y.~Du, B.~Dong, Z.~Chen, X.~Wang, Z.~Liu, P.~Gao, and S.~Li, ``{Efficient
		multi-user detection for uplink grant-free NOMA: Prior-information aided
		adaptive compressive sensing perspective},'' \emph{IEEE Journal on Selected
		Areas in Communications}, vol.~35, no.~12, pp. 2812--2828, 2017.
	
	\bibitem{wang2015compressive}
	B.~Wang, L.~Dai, Y.~Yuan, and Z.~Wang, ``{Compressive sensing based multi-user
		detection for uplink grant-free non-orthogonal multiple access},'' in
	\emph{IEEE 82nd Vehicular Technology Conference (VTC Fall)}, 2015.
	
	\bibitem{gao2016compressive}
	Z.~Gao, L.~Dai, Z.~Wang, S.~Chen, and L.~Hanzo, ``{Compressive-sensing-based
		multiuser detector for the large-scale SM-MIMO uplink},'' \emph{IEEE
		Transactions on Vehicular Technology}, vol.~65, no.~10, pp. 8725--8730, 2016.
	
	\bibitem{garcia2015low}
	A.~Garcia-Rodriguez and C.~Masouros, ``Low-complexity compressive sensing
	detection for spatial modulation in large-scale multiple access channels,''
	\emph{IEEE Transactions on Communications}, vol.~63, no.~7, pp. 2565--2579,
	2015.
	
	\bibitem{de2017random}
	E.~De~Carvalho, E.~Bj{\"o}rnson, J.~H. S{\o}rensen, E.~G. Larsson, and
	P.~Popovski, ``{Random Pilot and Data Access in Massive MIMO for Machine-Type
		Communications},'' \emph{IEEE Transactions on Wireless Communications},
	vol.~16, no.~12, pp. 7703--7717, 2017.
	
	\bibitem{liu2017massive}
	L.~Liu and W.~Yu, ``{Massive device connectivity with massive MIMO},'' in
	\emph{IEEE International Symposium on Information Theory (ISIT)}, 2017, pp.
	1072--1076.
	
	\bibitem{liu2017novel}
	W.~Liu, X.~Zhou, S.~Durrani, and P.~Popovski, ``A novel receiver design with
	joint coherent and non-coherent processing,'' \emph{IEEE Transactions on
		Communications}, vol.~65, no.~8, pp. 3479--3493, 2017.
	
	\bibitem{jing2016design}
	L.~Jing, E.~De~Carvalho, P.~Popovski, and A.~O. Martinez, ``Design and
	performance analysis of non-coherent detection systems with massive receiver
	arrays.'' \emph{IEEE Trans. Signal Processing}, vol.~64, no.~19, pp.
	5000--5010, 2016.
	
	\bibitem{zheng2002communication}
	L.~Zheng and D.~N.~C. Tse, ``Communication on the grassmann manifold: A
	geometric approach to the noncoherent multiple-antenna channel,'' \emph{IEEE
		Transactions on Information Theory}, vol.~48, no.~2, pp. 359--383, 2002.
	
	\bibitem{kim2011belief}
	\BIBentryALTinterwordspacing
	J.~Kim, W.~Chang, B.~Jung, D.~Baron, and J.~C. Ye, ``Belief propagation for
	joint sparse recovery,'' 2011. [Online]. Available:
	\url{https://arxiv.org/abs/1102.3289}
	\BIBentrySTDinterwordspacing
	
	\bibitem{ziniel2013efficient}
	J.~Ziniel and P.~Schniter, ``Efficient high-dimensional inference in the
	multiple measurement vector problem,'' \emph{IEEE Transactions on Signal
		Processing}, vol.~61, no.~2, pp. 340--354, 2013.
	
	\bibitem{rangan2016vector}
	\BIBentryALTinterwordspacing
	S.~Rangan, P.~Schniter, and A.~Fletcher, ``Vector approximate message
	passing,'' 2016. [Online]. Available: \url{https://arxiv.org/abs/1610.03082}
	\BIBentrySTDinterwordspacing
	
	\bibitem{donoho2009message}
	D.~L. Donoho, A.~Maleki, and A.~Montanari, ``Message-passing algorithms for
	compressed sensing,'' \emph{Proceedings of the National Academy of Sciences},
	vol. 106, no.~45, pp. 18\,914--18\,919, 2009.
	
	\bibitem{rangan2011generalized}
	S.~Rangan, ``Generalized approximate message passing for estimation with random
	linear mixing,'' in \emph{IEEE International Symposium on Information Theory
		(ISIT)}, 2011, pp. 2168--2172.
	
	\bibitem{bayati2011dynamics}
	M.~Bayati and A.~Montanari, ``The dynamics of message passing on dense graphs,
	with applications to compressed sensing,'' \emph{IEEE Transactions on
		Information Theory}, vol.~57, no.~2, pp. 764--785, 2011.
	
	\bibitem{liu2017massive2}
	L.~Liu and W.~Yu, ``{Massive connectivity with massive MIMO-part I: Device
		activity detection and channel estimation},'' \emph{arXiv preprint
		arXiv:1706.06438}, 2017.
	
	\bibitem{massivemimobook}
	\BIBentryALTinterwordspacing
	E.~Bj\"{o}rnson, J.~Hoydis, and L.~Sanguinetti, ``Massive {MIMO} networks:
	{Spectral}, energy, and hardware efficiency,'' \emph{Foundations and
		Trends{\textregistered} in Signal Processing}, vol.~11, no. 3-4, pp.
	154--655, 2017. [Online]. Available:
	\url{http://dx.doi.org/10.1561/2000000093}
	\BIBentrySTDinterwordspacing
	
	\bibitem{tropp2006algorithms}
	J.~A. Tropp, A.~C. Gilbert, and M.~J. Strauss, ``{Algorithms for simultaneous
		sparse approximation. Part I: Greedy pursuit},'' \emph{Signal Processing},
	vol.~86, no.~3, pp. 572--588, 2006.
	
	\bibitem{polyanskiy2010channel}
	Y.~Polyanskiy, H.~V. Poor, and S.~Verd{\'u}, ``Channel coding rate in the
	finite block-length regime,'' \emph{IEEE Transactions on Information Theory},
	vol.~56, no.~5, pp. 2307--2359, 2010.
	
	\bibitem{bayati2015universality}
	M.~Bayati, M.~Lelarge, A.~Montanari \emph{et~al.}, ``Universality in polytope
	phase transitions and message passing algorithms,'' \emph{The Annals of
		Applied Probability}, vol.~25, no.~2, pp. 753--822, 2015.
	
	\bibitem{gautschi1998incomplete}
	W.~Gautschi, ``The incomplete gamma functions since tricomi,'' in \emph{In
		Tricomi's Ideas and Contemporary Applied Mathematics, Atti dei Convegni
		Lincei, n. 147, Accademia Nazionale dei Lincei}.\hskip 1em plus 0.5em minus
	0.4em\relax Citeseer, 1998.
	
	\bibitem{temme1979asymptotic}
	N.~Temme, ``The asymptotic expansion of the incomplete gamma functions,''
	\emph{SIAM Journal on Mathematical Analysis}, vol.~10, no.~4, pp. 757--766,
	1979.
	
\end{thebibliography}
\end{document}